\documentstyle[12pt]{article}
\begin{document}

\newcommand{\bqn}{\begin{eqnarray}}
\newcommand{\eqn}{\end{eqnarray}}
\newcommand{\beq}{\begin{equation}}
\newcommand{\eeq}{\end{equation}}
\newcommand{\Hh}{{\textstyle\frac 12}}
\newcommand{\Hf}{{\textstyle\frac 14}}
\newcommand{\mbf}[1]{\mbox{\boldmath$#1$}}
\newcommand{\nrmp}{|\mbox{\boldmath $p$}|}

\newcommand{\CA}{{\cal A}}
\newcommand{\CS}{{\cal S}}
\newcommand{\CF}{{\cal F}}
\newcommand{\CG}{{\cal G}}
\newcommand{\CH}{{\cal H}}
\newcommand{\CI}{{\cal I}}
\newcommand{\CT}{{\cal T}}
\newcommand{\CM}{{\cal M}}
\newcommand{\CN}{{\cal N}}
\newcommand{\CL}{{\cal L}}
\newcommand{\CK}{{\cal K}}
\newcommand{\CO}{{\cal O}}
\newcommand{\CP}{{\cal P}}
\newcommand{\CR}{{\cal R}}
\newcommand{\CV}{{\cal V}}

\newcommand{\ga}{\alpha}
\newcommand{\gb}{\beta}
\newcommand{\gc}{\gamma}
\newcommand{\gd}{\delta}
\newcommand{\gk}{\kappa}
\newcommand{\gve}{\varepsilon}
\newcommand{\gl}{\lambda}
\newcommand{\gs}{\sigma}
\newcommand{\go}{\omega}
\newcommand{\gth}{\theta}
\newcommand{\gD}{\Delta}
\newcommand{\gL}{\Lambda}
\newcommand{\gO}{\Omega}
\newcommand{\Y}{{\Upsilon}}

\def\Spp{^3S_1^{++}}
\def\Dpp{^3D_1^{++}}
\def\Smm{^3S_1^{--}}
\def\Dmm{^3D_1^{--}}
\def\Pte{^3P_1^{e}}
\def\Pto{^3P_1^{o}}
\def\Pe{^1P_1^{e}}
\def\Po{^1P_1^{o}}

\begin{center}
{\Large \bf Magnetic moment of the deuteron as probe of \\
relativistic corrections}

\vskip 5mm
 S.G. Bondarenko, V.V.Burov\\
{\em{Laboratory of Theoretical Physics,
JINR,
141980 Dubna,Russia }}\\
 M.Beyer\\
{\em {Max Planck AG, Rostock University, 18055 Rostock, Germany}}\\
 S.M.Dorkin\\
{\em {Far East State University, 690000 Vladivostok, Russia}}\\

\end{center}

\begin{quote}
\small{\bf Abstract:}
  The calculation of the magnetic moment of the deuteron in the
  framework of the Bethe--Salpeter approach is performed. The relativistic
  corrections are calculated analytically and estimated numerically.
  It is shown that the main contributions are due to partial waves
  with positive energies and $P$--waves.  A comparison with the
  non-relativistic schemes of calculations including mesonic exchange
  currents is made.
\end{quote}

\section{Introduction}

The relativistic description of the deuteron has a long history. This
problem still remains present although non-relativistic schemes of
calculations are widely used to analyze reactions with the deuteron.
There are different approaches which take into account relativistic
effects.  They can be separated into three groups: (i)
the Bethe--Salpeter approach \cite{bethe}, (ii) a reduction of the
Bethe--Salpeter equation (quasi-potential, light front dynamics), and (iii)
prescriptional treatments.

The Bethe--Salpeter approach is the most consistent one. However, it
contains some difficulties that do not allow to establish a direct
link to the non-relativistic calculations -- e.g. the absence of a
non-relativistic reduction of arbitrary kernels or the problem to
interpret the abnormal parity states.

Non-relativistic calculations with mesonic exchange currents give a
reasonable description of the electromagnetic processes of the deuteron
(e.g. elastic scattering and electro disintegration). It is important
to note that those {\em mesonic exchange currents treated in a
  non--relativistic framework} are partially included in the above
mentioned relativistic calculations (i) and (ii) without additional assumptions.
This has been demonstrated e.g.  in ref. \cite{karm} for the
electro disintegration of the deuteron in the formalism of light front
dynamics.

Here, we present the investigation of a simple electromagnetic
property of the deuteron, the magnetic moment, in the framework of the
Bethe--Salpeter approach, and show the connection to the results of
a non-relativistic calculation.

\section{Basic definitions}

The main object of the Bethe--Salpeter approach is the vertex function
$\Gamma(P,p)$ which obeys the Bethe--Salpeter equation, here written within
the matrix formalism:
\beq
\left[ \Gamma(P,p) C \right] = i  \int \frac{d^4k}{(2\pi)^4}\; v(p,k)
\Gamma^{(1)} {S^{(1)}}(\frac{P}{2}+k) \left[ \Gamma(P,k) C \right]
{{\tilde S}^{(2)}}(\frac{P}{2}-k)\; \Gamma^{(2)},
\label{gammaeq}
\eeq
where $C$ is charge conjugation matrix, $v(p,k) \Gamma^{(1)}\otimes
\Gamma^{(2)}$ is the interaction
kernel, and  $S^{(i)}$, $\tilde S^{(i)}$ are Fermi propagators.
The vertex function for the deuteron can be decomposed into basis
states of definite values of orbital momentum $L$, spin $S$ and
$\rho$--spin $\rho$ \cite{tjon}, \cite{kubis}.
The general form in the rest frame is given by
\beq
\Gamma_{M}(P,p)=\sum \limits_{\alpha} \;
g_{\alpha} (p_0,|\mbf{p}|)\;
{\Gamma}_M^{\alpha}(\mbf{p}),
\label{reldp}
\eeq
where $g_{\alpha}$ are radial functions, and ${\Gamma}_M^{\alpha}(\mbf{p})$ are spin -- angular momentum parts.

In order  to exhibit the $\rho$-spin dependence,
the spin -- angular momentum
functions ${\Gamma}^\alpha_M(\mbox {\boldmath$p$})$ may be
replaced via
 ${\Gamma}^\alpha_M(\mbox {\boldmath$p$}) \equiv
 {\Gamma}^{\tilde\alpha,\,\rho_1\rho_2}_M(\mbf{p})$, where
\bqn
{\Gamma}^{\tilde\alpha,\, ++}_M(\mbox {\boldmath$p$}) &=
&\frac{\hat p_2 + m}{\sqrt{2E(m+E)}}\;
\frac{1+\gamma_0}{2}\;
{\tilde \Gamma}^{\tilde\alpha}_M(\mbox {\boldmath$p$})\;
\frac{\hat p_1 - m}{\sqrt{2E(m+E)}},
\nonumber\\
{\Gamma}^{\tilde\alpha,\, --}_M(\mbox {\boldmath$p$}) &=
&\frac{\hat p_1 - m}{\sqrt{2E(m+E)}}\;
\frac{-1+\gamma_0}{2}\;
{\tilde \Gamma}^{\tilde\alpha}_M(\mbox {\boldmath$p$})\;
\frac{\hat p_2 + m}{\sqrt{2E(m+E)}},
\label{gf}\\
{\Gamma}^{\tilde\alpha,\, +-}_M(\mbox {\boldmath$p$}) &=
&\frac{\hat p_2 + m}{\sqrt{2E(m+E)}}\;
\frac{1+\gamma_0}{2}\;
{\tilde \Gamma}^{\tilde\alpha}_M(\mbox {\boldmath$p$})\;
\frac{\hat p_2 + m}{\sqrt{2E(m+E)}},
\nonumber\\
{\Gamma}^{\tilde\alpha,\, -+}_M(\mbox {\boldmath$p$}) &=
&\frac{\hat p_1 - m}{\sqrt{2E(m+E)}}\;
\frac{1-\gamma_0}{2}\;
{\tilde \Gamma}^{\tilde\alpha}_M(\mbox {\boldmath$p$})\;
\frac{\hat p_1 - m}{\sqrt{2E(m+E)}},
\nonumber
\eqn
with $\tilde\alpha\in\{L,S,J\}$, $m$ the nucleon mass, $\hat p =
\gamma_\mu p^\mu$, and $\tilde
\Gamma_M^{\tilde\alpha}$ given by
\[
\begin{array}{cc}
\tilde\alpha&{\sqrt{8\pi}\;\;\tilde \Gamma}_M^{\tilde\alpha}\\[1ex]
\hline
^3S_1&{\hat \xi_M}\\
^3D_1& -\frac{1}{\sqrt{2}}
\left[ {\hat \xi_M}+\frac{3}{2}
({\hat p_1}-{\hat p_2})(p\xi_M)\nrmp^{-2}\right]\\
^3P_1& \sqrt{\frac{3}{2}}
\left[ \frac{1}{2} {\hat \xi_M}({\hat p_1}-{\hat p_2})
-(p\xi_M) \right]\nrmp^{-1}\\
^1P_1&\sqrt{3} (p\xi_M)\nrmp^{-1}\\
\end{array}
\]

There are eight states in the deuteron channel (instead of two in
the non-relativistic case), viz. $\Spp$, $\Dpp$, $\Smm$, $\Dmm$, $\Pte$,
$\Pto$, $\Pe$, $\Po$ (notation: $^{2S+1} L _J^{\rho_1 \rho_2 }$).
The normalization condition for this functions can be written as:
\bqn
P_{+}+P_{-}=1,\qquad\qquad
P_+ &=& P_{^3S_1^{++}} + P_{^3D_1^{++}}, \nonumber \\
P_- &=& P_{^3S_1^{--}} + P_{^3D_1^{--}} + P_{^3P_1^{e}} +
P_{^1P_1^{o}} + P_{^3P_1^{e}} + P_{^1P_1^{e}},
\eqn
introducing pseudo-probabilities $P_\alpha$ that are negative
for the states $\Smm$, $\Dmm$, $\Pte$,
$\Pto$, $\Pe$, $\Po$, and positive for $\Spp$, $\Dpp$
\cite{tjon}. The calculation with realistic vertex functions give the
following values:
\begin{center}
\begin{tabular}{cccccc}
    &$\Dpp$ & $\Dmm$         & $\Pte +\Pto$        & $\Pe +\Po$& \\
\hline
[\%]& $4.8$ & $-6\cdot 10^{-4}$  & $-0.88 \cdot 10^{-2}$& $-2.5 \cdot
10^{-2}$&\protect{\cite{tjon}}\\[0ex]
[\%]& $5.1$ & $-3.4\cdot 10^{-4}$& $-9 \cdot 10^{-2}$  & $-2.4 \cdot
10^{-2}$&\protect{\cite{kazakov}}
\end{tabular}
\end{center}
It is obvious that the main contribution to the normalization is due
to the states with positive energies, and the contribution of the
$P$--states is larger than that of the negative energies states by at
least one order of magnitude.

\section{The calculation of the magnetic moment}

The matrix element of the electromagnetic current used to evaluate the
magnetic moment of the deuteron is given by
\bqn
\lefteqn{
\langle P^{\prime} M^{\prime} | J_{x} | P M \rangle =
\frac{ie}{2\pi^2 M} \int d^4p } \nonumber\\
&&Tr \biggl\{ {\bar \Gamma_{M^{\prime}}}(P^{\prime},p^{\prime})
S^{(1)}(\frac{P^{\prime}}{2}+p^{\prime})
\bigl[ \gamma_1 F_1^{(s)}(q^2) -
\frac{\gamma_1 {\hat q} - {\hat q} \gamma_1}{4m} F_2^{(s)}(q^2) \bigr]
\Psi_{M}(P,p)
\biggr\},
\label{emcbsiso}
\eqn
where
$\Psi_{M}(P,p) = S^{(1)}(\frac{P}{2}+p) \Gamma_{M}(P,p)
{\tilde S^{(2)}}(\frac{P}{2}-p)$,  the
isoscalar  form factors of the nucleons are given by
$F_{1,2}^{s}(q^2)$, and  $p^{\prime}=p+q/2$,
$P^{\prime}=P+q$. The magnetic moment  then is evaluated via
\beq
\mu_D = \frac{1}{e} \frac{m}{M} \sqrt{2} \lim_{\eta \to 0}
\frac{\langle M^{\prime}=+1 | J_{x} | M=0 \rangle}{\sqrt{\eta}\sqrt{1+\eta}},
\label{magmom}
\eeq
with $\eta=-q^2/4 M^2$.
Since the integral (\ref{emcbsiso}) vanishes at $q^2=0$, we expand the
integrand of (\ref{emcbsiso}) in powers of $\sqrt{\eta}$. This is done
in the Breit system. To this end the vertex functions need to be
transformed from the rest system to the Breit system using the following
formulae.
\begin{eqnarray}
\Gamma_{M}(P^{(B)},p^{(B)}) &=& \Lambda(P^{(B)})\; \Gamma_{M}(P^{(0)},p^{(0)})\;
\Lambda^{-1}(P^{(B)}), \nonumber\\
\Gamma_{M}(P^{\prime \ (B)},p^{\prime \ (B)}) &=&
\Lambda^{-1}(P^{(B)})\; \Gamma_{M}(P^{(0)},p^{\prime \ (0)})\;
\Lambda(P^{(B)}),
\nonumber
\label{transff}
\end{eqnarray}
where $\Lambda(P^{(B)})=
( M + {\hat P}^{(B)} \gamma_0)/\sqrt{2M(E^{(B)}+M)}$,
and the vectors $P^{(B)}$, $P^{\prime \ (B)}$, $p^{(B)}$, $p^{\prime \
  (B)}$ are connected
with the respective vectors in the rest system by
\begin{eqnarray}
P^{(B)} = {\cal L} P^{(0)}, \quad p^{(B)} = {\cal L} p^{(0)}, \quad
P^{\prime \ (B)} = {\cal L}^{-1} P^{(0)}, \quad
p^{\prime \ (B)} = {\cal L}^{-1} p^{\prime \ (0)}.
\label{transfv}
\end{eqnarray}
The Lorentz transformation matrix ${\cal L}$ is of the form
\begin{eqnarray}
{\cal L} = \left(
\begin{array}{cccc}
\sqrt{1+\eta} & 0 & 0 & -\sqrt{\eta} \\
0 & 1 & 0 & 0 \\
0 & 0 & 1 & 0 \\
-\sqrt{\eta} & 0 & 0 & \sqrt{1+\eta}
\end{array}
\right).
\label{mtrtransf}
\end{eqnarray}

From (\ref{transfv}) it follows that $p^{\prime \ (0)} =
{\cal L} p^{\prime \ (B)} = {\cal L} \bigl( p^{(B)} + \frac{1}{2}q^{B}) =
{\cal L}^{2} p^{(0)} + \frac{1}{2} {\cal L}
q^{(B)}$. To simplify notation the
vector $p^{(0)}$ will be denoted as
$p^{(0)} \equiv p=(p_0,p_x,p_y,p_z)$. The components of the vector
$p^{\prime \ (0)}\equiv p^{\prime}$
are then given by
\begin{eqnarray}
p_0^{\prime} &=& (1+2\eta)p_0 - 2\sqrt{\eta}\sqrt{1+\eta}p_z - M\eta,
\nonumber\\
p_x^{\prime} &=& p_x, \quad p_y^{\prime} = p_y, \\
p_z^{\prime} &=& (1+2\eta)p_z - 2\sqrt{\eta}\sqrt{1+\eta}p_0 +
M\sqrt{\eta}\sqrt{1+\eta}.
\nonumber
\label{pprimecm}
\end{eqnarray}

With the help of the transformations
 (\ref{transff}), the integral (\ref{emcbsiso})
may then be written as
\begin{eqnarray}
\lefteqn{
\langle P^{\prime} M^{\prime} | J_{x} | P M \rangle =
\frac{ie}{2\pi^2 M} \int d^4p
\;Tr \left\{
{\bar \Gamma_{M^{\prime}}}(P^{(0)},p^{\prime})
S^{(1)}(\frac{P^{(0)}}{2}+p^{\prime})
\times\right.}
\nonumber\\
&&\left.\times
\Lambda(P^{(B)})
\bigl[ \gamma_1 F_1^{(s)}(q^2) -
\frac{\gamma_1 {\hat q} - {\hat q} \gamma_1}{4m} F_2^{(s)}(q^2) \bigr]
\Lambda(P^{(B)})
\Psi_{M}(P^{(0)},p)
{[\Lambda^{-1}(P^{(B)})]}^{2}
\right\},
\label{emcbscm}
\end{eqnarray}
where the wave function $\Psi_{M}(P^{(0)},p)$ and the vertex function
${\bar \Gamma_{M^{\prime}}}(P^{(0)},p^{\prime})$ are taken in
the deuteron rest frame.

It is obvious from (\ref{emcbscm}) that the sources of $\sqrt{\eta}$
terms may appear in
(i) the matrix $\Lambda(P^{(B)})$, (ii) the propagator
$S^{(1)}(\frac{1}{2}P^{(0)}+p^{\prime})$, and (iii) the vertex function
${\bar \Gamma_{M^{\prime}}}(P^{(0)},p^{\prime})$. In detail this reads
for the matrix (i)
\begin{eqnarray}
\Lambda(P^{(B)})
\bigl[ \gamma_1 F_1^{(s)}(q^2) -
\frac{\gamma_1 {\hat q} - {\hat q} \gamma_1}{4m} F_2^{(s)}(q^2) \bigr]
\Lambda(P^{(B)}) &=& \frac{1}{2} (\gamma_1 +
\sqrt{\eta}\gamma_1\gamma_3\gamma_0 - \frac{\kappa}{4m}(\gamma_1{\hat q}-
{\hat q}\gamma_1)), \nonumber\\
{[\Lambda^{-1}(P^{(B)})]}^{2}& =& 1 + \sqrt{\eta} \gamma_0 \gamma_3,
\label{lambdaexp}
\end{eqnarray}
and for the propagator (ii)
\beq
S^{(1)}(\frac{P^{(0)}}{2}+p^{\prime}) =
S^{(1)}(\frac{P^{(0)}}{2}+p) \Biggl[1 +
\sqrt{\eta} \frac{4Mp_z}{(\frac{P^{(0)}}{2}+p)^2-m^2} \Biggr] -
\sqrt{\eta} \frac{2p_z\gamma_0+(M-2p_0)\gamma_3}{(\frac{P^{(0)}}{2}+p)^2-m^2}.
\label{propexp}
\eeq

The corrections due to the vertex function (iii)
may be separated into  ($\alpha$) corrections due to the radial wave functions,
and ($\beta$) due to the spin -- angular momentum  function. This will be
demonstrated in the following using only one component
of the vertex function (\ref{reldp}), i.e.
\begin{eqnarray}
{\bar \Gamma_{M^{\prime}}}(P^{(0)},p^{\prime}) =  g(p_0^{\prime},
|\mbox{\boldmath$p$}^{\prime}|)\;
{\bar \Gamma_{M^{\prime}}(\mbox{\boldmath $p$}^{\prime})}.
\end{eqnarray}
For the radial part we get the following expansion
\begin{eqnarray}
g(p_0^{\prime},|\mbox{\boldmath$p$}^{\prime}|) = g(p_0,
|\mbox{\boldmath$p$}|) + \sqrt{\eta}p_z
\Bigl\{ -2 \frac{\partial}{\partial p_0} + \frac{M-2p_0}
{|\mbox{\boldmath$p$}|}
\frac{\partial}{\partial |\mbox{\boldmath$p$}|} \Bigr\}
g(p_0,|\mbox{\boldmath$p$}|).
\label{radexp}
\end{eqnarray}
The spin -- angular momentum part can be written (e.g. for  the
$^3D^{++}_1$ state) as
\begin{eqnarray}
{\bar \Gamma_{M^{\prime}}(\mbox{\boldmath $p$})} =
N(E^{\prime}) (m-{\hat p}_1^{\prime})
{\tilde \Gamma}(\mbox{\boldmath $p$}^{\prime}, \mbox{\boldmath $\xi)$}
(m+{\hat p}_2^{\prime}),
\label{spinoexp}
\end{eqnarray}
where
\begin{eqnarray}
{\hat p}_1^{\prime} &=&  {\hat p}_1 + \sqrt{\eta}(M-2p_0)
[\frac{p_z}{E}\gamma_0-\gamma_3], \nonumber\\
{\hat p}_2^{\prime} &=&  {\hat p}_2 + \sqrt{\eta}(M-2p_0)
[\frac{p_z}{E}\gamma_0+\gamma_3], \label{dsm}\\
N(E^{\prime})&=&\frac{1}{2E^{\prime}(m+E^{\prime})}=
\frac{1}{2E(m+E)}\Bigl(1-\sqrt{\eta}\frac{M-2p_0}{E^2(m+E)}(m+2E)p_z\Bigr),
\label{vecexp}
\end{eqnarray}
and corrections to  ${\tilde \Gamma}(\mbox{\boldmath $p$}^{\prime},
\mbox{\boldmath $\xi)$}$ can be calculated using (\ref{dsm})--(\ref{vecexp}).

\section{Results of calculations}

The general formula for magnetic moment can be written  as
\begin{eqnarray}
\mu&=&\mu_{+} + \mu_{1-} + \mu_{2-}
\nonumber\\
\mu_+&=&(\mu_p+\mu_n)(P_{\Spp}+P_{\Dpp}) -
\frac{3}{2}(\mu_p+\mu_n-\frac{1}{2})P_{\Dpp} + R_{+},\nonumber\\
\mu_{2-}&=&-(\mu_p+\mu_n)P_{\Smm}+P_{\Smm}+
\frac{1}{2}(\mu_p+\mu_n)P_{\Dmm} - \frac{5}{4}P_{\Dmm}+
R_{2-},
\nonumber\\
\mu_{1-}&=&-\frac{1}{2}(\mu_p+\mu_n)(P_{^3P_1^{e}}+P_{^3P_1^{o}})-
\frac{1}{4}(P_{^3P_1^{e}}+P_{^3P_1^{o}}) - \frac{1}{2}(P_{^1P_1^{e}}+
P_{^1P_1^{o}}) + R_{1-},
\nonumber
\end{eqnarray}
where $R_{a}$ are the relativistic correction terms, viz.
\begin{eqnarray}
R_{+} &=& -\frac{1}{3}(\mu_p+\mu_n) H_1^{\Spp} - \frac{1}{2} H_2^{\Spp}
- \frac{m}{M} H_3^{\Spp} - \left(1-\frac{2m}{M}\right) G_1^{\Spp} -
\nonumber\\
&&-\frac{1}{6}(\mu_p+\mu_n-3) H_1^{\Dpp} - \frac{1}{2} H_2^{\Dpp} -
\frac{m}{M} H_3^{\Dpp} + \left(1-\frac{2m}{M}\right) G_2^{\Dpp} +
\nonumber\\
&&+\frac{\sqrt{2}}{3} \left[\mu_p+\mu_n-(1+\frac{m}{M})\right] H_1^{\Spp,\Dpp},
\nonumber\\
R_{2-} &=& -\frac{1}{3}\left[\mu_p+\mu_n-(1-\frac{2m}{M})\right]
H_1^{\Smm} - \frac{m}{M} H_2^{\Smm} + \frac{m}{M} H_3^{\Smm} -
\nonumber\\
&&-\frac{1}{6}\left[\mu_p+\mu_n-1+\frac{4m}{M}\right] H_1^{\Dmm} -
\frac{m}{M} H_2^{\Dmm} + \frac{m}{M} H_3^{\Dpp} + \frac{3}{4}
(1-\frac{2m}{M}) P_{\Dmm} +
\nonumber\\
&&+\frac{\sqrt{2}}{3} \left[\mu_p+\mu_n-(1-\frac{m}{M})\right] H_1^{\Smm,\Dmm},
\nonumber\\
R_{1-} &=& \frac{1}{2}\left(1-\frac{2m}{M}\right)
\left[\mu_p+\mu_n+\frac{1}{2}\right] \left( P_{\Pte} + P_{\Pto} \right) -
\left[\mu_p+\mu_n+1\right] H_4^{\Pte,\Pto} +
\nonumber\\
&& + \frac{1}{2}\left(1-\frac{2m}{M}\right)
\left( P_{\Pe} + P_{\Po} \right) - 2 H_4^{\Pe,\Po} +
\nonumber\\
&& + G_4^{\Pte,\Pe} + G_4^{\Pto,\Po} + G_5^{\Pte,\Po} + G_6^{\Pto,\Pe},
\nonumber
\end{eqnarray}
and $H_{i}^{\alpha,\alpha^{\prime}} (H_{i}^{\alpha,\alpha} \equiv
H_{i}^{\alpha})$ and $G_{i}^{\alpha}$
are integrals of the form
$\frac{-i}{4 \pi^2 M} \int dp_0 |\mbox{\boldmath$p$}|^2 d|\mbox{\boldmath$p$}|$
with the integrands given by (compare with \cite{honzava})
\begin{eqnarray}
H_1^{\alpha,\alpha^{\prime}} &=&
\Bigl(1-\frac{m}{E}\Bigr)
\Bigl[\phi_{\alpha}(p_0,|\mbox{\boldmath$p$}|)
\phi_{\alpha^{\prime}}(p_0,|\mbox{\boldmath$p$}|)\Bigr]
\times
\Biggl\{
\begin{array}{l}
(E-M/2), \mbox{ for } \alpha(\alpha^{\prime})=\Spp,\Dpp, \\
(E+M/2), \mbox{ for } \alpha(\alpha^{\prime})=\Smm,\Dmm
\end{array}
\nonumber\\
H_2^{\alpha}& =&
\Bigl( \frac{1}{E}(E-\frac{M}{2})\Bigr)
\Bigl[\phi_{\alpha}(p_0,|\mbox{\boldmath$p$}|)\Bigr]^2
\times
\Biggl\{
\begin{array}{l}
(E-M/2), \mbox{ for } \alpha=\Spp,\Dpp, \\
(E+M/2), \mbox{ for } \alpha=\Smm,\Dmm
\end{array}
\nonumber\\
H_3^{\alpha} &=&
\Bigl( \frac{p_0^2}{E} \Bigr)
\Bigl[\phi_{\alpha}(p_0,|\mbox{\boldmath$p$}|)\Bigr]^2, \qquad
H_4^{\alpha,\alpha^{\prime}} =
\Bigl( \frac{m}{E}p_0 \Bigr)
\Bigl[\phi_{\alpha}(p_0,|\mbox{\boldmath$p$}|)
\phi_{\alpha^{\prime}}(p_0,|\mbox{\boldmath$p$}|)\Bigr]
\nonumber\\
G_1^{\Spp} &=&
\frac{2E+4m+3M}{12E} \Bigl(E-\frac{M}{2}\Bigr)
\Bigl[\phi_{\Spp}(p_0,|\mbox{\boldmath$p$}|)\Bigr]^2
\nonumber\\
G_2^{\Dpp} &=&
\frac{E-4m+3M}{12E} \Bigl(E-\frac{M}{2}\Bigr)
\Bigl[\phi_{\Dpp}(p_0,|\mbox{\boldmath$p$}|)\Bigr]^2
\nonumber\\
G_4^{\alpha,\alpha^{\prime}} &=&
\Bigl( -\frac{2\sqrt{2}m^2}{ME} p_0 \Bigr)
\Bigl[\phi_{\alpha}(p_0,|\mbox{\boldmath$p$}|)
\phi_{\alpha^{\prime}}(p_0,|\mbox{\boldmath$p$}|)\Bigr]
\nonumber\\
G_5^{\Pte,\Po} &=&
\Bigl( (\mu_p +\mu_n) \frac{\sqrt{2}M}{2} -
\frac{\sqrt{2}M}{2} (1-\frac{4m^2}{M^2}) \Bigr)
\Bigl[ \phi_{^3 P_1^{e}}(p_0,|\mbox{\boldmath$p$}|)
\phi_{^1 P_1^{o}}(p_0,|\mbox{\boldmath$p$}|) \Bigr]
\nonumber\\
G_6^{\Pto,\Pe} &=&
\Bigl( (\mu_p +\mu_n) \frac{\sqrt{2}M}{2} -
\frac{\sqrt{2}M}{2}(1-\frac{m^2}{E^2}(1+\frac{4p_0^2}{M^2})) \Bigr)
\Bigl[ \phi_{^3 P_1^{o}}(p_0,|\mbox{\boldmath$p$}|)
\phi_{^1 P_1^{e}}(p_0,|\mbox{\boldmath$p$}|) \Bigr]
\nonumber
\end{eqnarray}

The relativistic corrections  $R_a$ can be estimated numerically for
a separable model \cite{rupp}.
They are smaller than the dominant terms by at least one order of magnitude.
Taking into account different orders of pseudo-probabilities one gets
\begin{eqnarray}
\mu_d&=&\mu_{NR} + \Delta \mu
\label{main}\\
\mu_{NR}&=&
(\mu_p+\mu_n)+\frac{3}{2}(\mu_p+\mu_n-\frac{1}{2})P_{\Dpp} ,\nonumber\\
\Delta \mu&=&R_+ + \Delta \mu_{1-}
\nonumber\\
\Delta \mu_{1-}&=&-\frac{3}{2}(\mu_p+\mu_n)(P_{^3P_1^{e}}+P_{^3P_1^{o}})-
(\mu_p+\mu_n) (P_{^1P_1^{e}}+P_{^1P_1^{o}})\nonumber\\&&
-\frac{1}{4}(P_{^3P_1^{e}}+P_{^3P_1^{o}}) - \frac{1}{2}(P_{^1P_1^{e}}+
P_{^1P_1^{o}}),
\nonumber
\end{eqnarray}

The contribution $R_+$ is negative and $\Delta \mu_{1-}$ is positive.
An estimation gives the following results
\beq
R_+ / \mu_{NR}=-(5.8 - 11)\times 10^{-2}\; \%, \qquad
\Delta_ {\mu{1-}} / \mu_{NR}=0.2\; \%
\label{res}
\eeq

\section{Conclusion}

We have shown that  the expression
for the magnetic moment in the Bethe--Salpeter approach
can be written in a form closer to non-relativistic calculations. The
additional terms in  equation (\ref{main}) can be considered as
relativistic corrections
to the non-relativistic formula.

The experimental value of the magnetic moment is known with a high accuracy:
$\mu_{exp} = 0.857406 (1) $. The non-relativistic value reflects
only the $D$-state probability. Whereas in the relativistic corrections
$P$-states  play  the dominant role.

The magnitude  of the corrections can be compared
with the contributions of mesonic
exchange currents to the magnetic moment as extracted from ref. \cite{burov}.
The main contribution is due to the pair term, which leads to
$\Delta \mu /\mu_{NR}=0.21-0.22\%$
for different forms of the Bonn potential.
The same size of this  correction as compared to (\ref{res})
may be considered as an indication that both corrections
are of  the same physical origin.

This work has been partially supported by the Deutsche Akademische 
Austauschdienst.

\end{document}